\documentclass[twocolumn]{aastex7}

\newcommand{\abund}[1]{$\log N({\rm #1})/N({\rm H})$}

\newcommand{\figref}[1]{Fig.~\ref{#1}}
\newcommand{\h}{$^{\mathrm h}$}
\newcommand{\ie}{i.e.}
\newcommand{\m}{$^{\mathrm m}$}

\newcommand{\hone}{\ion{H}{1}}

\newcommand{\heone}{\ion{He}{1}}
\newcommand{\hetwo}{\ion{He}{2}}

\newcommand{\cthree}{\ion{C}{3}}
\newcommand{\cfour}{\ion{C}{4}}

\newcommand{\ntwo}{\ion{N}{2}}
\newcommand{\nthree}{\ion{N}{3}}
\newcommand{\nfour}{\ion{N}{4}}
\newcommand{\nfive}{\ion{N}{5}}

\newcommand{\otwo}{\ion{O}{2}}
\newcommand{\othree}{\ion{O}{3}}
\newcommand{\ofour}{\ion{O}{4}}
\newcommand{\ofive}{\ion{O}{5}}

\newcommand{\netwo}{\ion{Ne}{2}}

\newcommand{\mgtwo}{\ion{Mg}{2}}

\newcommand{\althree}{\ion{Al}{3}}

\newcommand{\sithree}{\ion{Si}{3}}
\newcommand{\sifour}{\ion{Si}{4}}

\newcommand{\pfour}{\ion{P}{4}}
\newcommand{\pfive}{\ion{P}{5}}

\newcommand{\sthree}{\ion{S}{3}}
\newcommand{\sfour}{\ion{S}{4}}
\newcommand{\sfive}{\ion{S}{5}}
\newcommand{\ssix}{\ion{S}{6}}

\newcommand{\clfour}{\ion{Cl}{4}}

\newcommand{\fethree}{\ion{Fe}{3}}
\newcommand{\fefour}{\ion{Fe}{4}}

\newcommand{\nifour}{\ion{Ni}{4}}

\newcommand{\gefour}{\ion{Ge}{4}}

\newcommand{\ebv}{$E(B-V)$\/}
\newcommand{\f}{{\it f}\/}
\newcommand{\flux}{erg cm$^{-2}$ s$^{-1}$  \AA$^{-1}$}

\newcommand{\kms}{km s$^{-1}$}
\newcommand{\logg}{$\log g$}

\newcommand{\msun}{$M_{\sun}$}

\newcommand{\teff}{$T_{\rm eff}$}

\graphicspath{{./}{Figures/}}

\begin{document}

\title{Observations of the Halo Star HD 177566}

\author[orcid=0000-0001-9184-4716,gname='William V.',sname='Dixon']{William V. Dixon}
\affiliation{Space Telescope Science Institute}
\email[show]{dixon@stsci.edu}  

\author[orcid=0000-0001-7653-0882,gname=Pierre, sname='Chayer']{Pierre Chayer}
\affiliation{Space Telescope Science Institute}
\email{chayer@stsci.edu}



\begin{abstract}

We have analyzed archival FUV and optical spectra of the hot halo star HD~177566.  The star has an effective temperature $T_{\rm eff} = 33{,}000 \pm 1000$ K, surface gravity $\log g = 3.79 \pm 0.11$, and helium abundance $\log N({\rm He})/N({\rm H}) = -0.86 \pm 0.05$.  Abundances of 13 additional elements are consistent with those of other halo stars, save for carbon, which is underabundant by about 1 dex.  The low-order hydrogen Balmer lines are not well reproduced by our models.  The diffuse lines of \ion{He}{1} are often broader than predicted, but the use of more recent line-broadening parameters significantly improves the fit.  Scaling our best-fit model to the star's optical and near-IR magnitudes yields an extinction $E(B-V) = 0.095 \pm 0.005$, consistent with literature values, but the resulting model underpredicts the star's FUV flux by a factor of two.  The star's effective temperature and luminosity ($\log L / L_{\sun} = 3.50 \pm 0.08$) place it on the post-AGB evolutionary tracks of a star that evolved from the red horizontal branch.  Its low carbon abundance, $\log N({\rm C})/N({\rm O}) = -2.18 \pm 0.21$, indicates that it did not experience significant third dredge-up while ascending the AGB.

\end{abstract}


\keywords{\uat{Spectroscopy}{1558} --- \uat{Stellar abundances}{1577} --- \uat{Stellar atmospheres}{1584} --- \uat{Stellar evolution}{1599}}


\section{Introduction} 

In the early 1980's, the presence of apparently normal O- and B-type Population I stars at high galactic latitudes presented a challenge to stellar astronomy.  Since these stars could not have formed in the low-density gas of the halo, they must have been ejected from the disk at velocities as high as several hundred \kms\ \citep{House:Kilkenny:1978}.  One example is HD~177566, a 10th-magnitude star \citep[$V = 10.20$;][]{Hill:1970} with spectral type B5/7 IB \citep{Houk:1978} and radial velocity $V_{\rm rad} = -134$ \kms\ \citep{Gontcharov:2006}.  \citet{Keenan:Dufton:1983} used Stromgren $| C_1 |$ and H$\beta$ photometry to estimate the star's effective temperature and surface gravity; they derived \teff\ = 31{,}000 K and \logg\ = 3.5 and concluded that it is a main-sequence star at a distance of some 12 kpc.  A decade later, it was clear that these early-type, high-latitude stars are not massive main-sequence stars, but low-mass post-asymptotic giant branch (post-AGB) stars considerably closer to the Sun.  From an LTE model-atmosphere analysis of its optical and ultraviolet spectra, \citet{Kendall:1994} derived photospheric parameters of \teff\ = $30{,}000 \pm 1000$ K and \logg\ = $3.8 \pm 0.2$ for HD~177566.  They found that the star is underabundant in heavy elements (relative to the sun) by roughly $-1.3$ dex and concluded that it is a post-AGB star with mass $M/M_{\sun} \approx 0.55$.  

While the general nature of early-type stars in the halo is now understood, HD~177566 remains something of a mystery.  \citet{Mello:2012} observed ten hot stars thought to be post-AGB objects.  Applying non-LTE models to a high-resolution optical spectrum of HD~177566, the authors derived photospheric parameters of \teff\ = $30{,}900 \pm 300$ K and \logg\ = 3.83, similar to those of previous observers.  Their spectrum exhibits no emission lines, but the cores of its H$\alpha$ and H$\beta$ profiles are flattened, which may indicate the presence of weak emission.  They measured the star's He, N, O, Mg, and Si abundances.  Unlike the nine other stars in their sample, which have solar metallicities, HD~177566 is metal poor, with a metal mass fraction of $Z = 0.001$. The authors suggest several possible evolutionary scenarios.  In order of descending zero-age horizontal branch (ZAHB) mass, HD~177566 may be a post-AGB star that fully ascended the AGB, a post-early AGB star that left the AGB before the onset of thermal pulsing, or an AGB manqu\'{e} star that is evolving directly from the HB without ascending the AGB.  A more exotic possibility is that the star is an O-type hot subdwarf that is evolving from the AGB toward the helium main sequence.

\citet{Partha:2020} brought some clarity to the situation by using the stellar distance from Gaia DR2 to derive the luminosity of HD~177566.  With a distance of 1,436 $\pm$ 219 pc, the star has a luminosity $\log (L/L_{\sun}) = 2.59 \pm 0.14$, considerably fainter than the post-AGB evolutionary tracks of \citet{Miller_Bertolami:2016} and closer to the AGB manqu\'{e} tracks of \citet{Dorman:1993}.  They suggest that it is a post-early AGB or AGB manqu\'{e} star.  The star's kinematics are inconsistent with Galactic rotation; its high velocity relative to the Sun suggests that it belongs to the Galactic halo.

To better understand the nature of HD~177566, we have analyzed archival UV and optical spectra of the star.  In Section \ref{sec:observations}, we present our data.  In Section \ref{sec:analysis}, we present our atmospheric models and use them to derive stellar parameters and abundances.  In Section \ref{sec:discussion}, we discuss our results and consider the evolutionary history of the star.  We summarize our conclusions in Section \ref{sec:conclusion}.

\begin{deluxetable*}{lcccccl}
\tablecaption{Summary of Observations \label{tab:log_obs}}
\tablehead{
\colhead{Instrument} & \colhead{Wavelength} & \colhead{$R\equiv\lambda/\Delta\lambda$} & \colhead{Exp. Time} & \colhead{Obs.\ Date} & \colhead{Data ID} & \colhead{P.I.} \\
& \colhead{Range (\AA)} & & \colhead{(s)}
}
\startdata
FUSE & \phn905--1187 & 20,000      &   3387 & 2000 Aug 29 & P1017001 & Sembach \\
           &                &             &   7764 & 2000 Sep 18 & P1017003 & Sembach \\
           &                &             &   2576 & 2003 Sep 21 & M1051301 & Dupuis \\
STIS/E140M    & 1140--1729 & 45,800 & 2043 & 2014 Aug 28 & OC8Y04010 & Fox \\
IUE             & 1150--1975 &  \phn8,400    & 4200 & 1979 Mar 22 & SWP04726 & Wilson \\
                   &                    &              &  3600 & 1983 Feb 26 & SWP19328 & Pettini \\
UVES-Blue & 3290--4520 & 65,030 & 1080 & 2013 Sep 19 & 092.C--0173(A) & Smoker \\
UVES-Red & 4790--6800 & 74,450 \\
\enddata
\end{deluxetable*}

\section{Observations and Data Reduction}\label{sec:observations}

The star has been observed with the Far Ultraviolet Spectroscopic Explorer (FUSE), the Space Telescope Imaging Spectrograph (STIS) aboard the {\em Hubble Space Telescope}, the International Ultraviolet Explorer (IUE), and the Ultraviolet and Visual Echelle Spectrograph (UVES) on the European Southern Observatory's VLT.  Observational details are presented in Table \ref{tab:log_obs}.  The FUV data used in this work are available at \dataset[10.17909/m3rp-8x48]{https://doi.org/10.17909/m3rp-8x48}.

\subsection{FUSE Spectroscopy}

FUSE provides medium-resolution spectroscopy from 1187 \AA\ to the Lyman limit \citep{Moos:2000, Sahnow:2000}.  The star HD~177566 was observed through the $30\arcsec \times 30\arcsec$ LWRS aperture.  The data were reduced using v3.2.2 of CalFUSE, the standard data-reduction pipeline software \citep{Dixon:2007}, and retrieved from the Mikulski Archive for Space Telescopes (MAST).  For each FUSE channel, the extracted spectra from all exposures are shifted to a common wavelength scale, weighted by exposure time, and combined into a single file.  Because each channel has a unique line-spread function, we do not combine the spectra from multiple channels, instead using only the spectrum from the channel with the highest signal-to-noise ratio (S/N).  The S/N of the resulting spectrum is $\sim 100$ per 0.05 \AA\ resolution element in the SiC band (900--1000 \AA) and $\sim 150$ in the LiF band (1000--1187 \AA).  The FUSE wavelength calibration is reasonably accurate, but small offsets of the target within the LWRS aperture can introduce a zero-point offset in the wavelength scale.   To place the data on an absolute wavelength scale, we shift the spectrum so that the velocities of its interstellar lines match those of the STIS spectrum, which are heliocentric.  

The FUSE LWRS aperture is relatively large, so we should confirm that there are no other FUV sources in the field.  HD~177566 was not observed by GALEX, but the Gaia DR3 catalog lists eleven objects within 25\arcsec\ of the star \citep{Gaia_Mission:2016, Gaia_DR3:2023}.  The brightest, at G = 14.4 mag, has an estimated temperature of \teff\ = 5900 K, too cool to produce significant FUV emission.  We conclude that the FUSE spectrum is not contaminated by other sources. 

{
We can check the photometric accuracy of our data by looking at the jitter files for each exposure.  They show that the pointing was stable to within $\pm 1\arcsec$ for most exposures and to within $\pm 2\arcsec$ for all exposures, sufficient to keep the star within the LWRS aperture.
}

\subsection{STIS Spectroscopy}

The design and construction of STIS are described by \citet{Woodgate:STIS:1998}, and its on-orbit performance is discussed by \citet{Kimble:STIS:1998}.  HD~177566 was observed using the STIS far-ultraviolet MAMA detector with the E140M echelle grating and the $0\farcs2 \times 0\farcs06$ aperture.  We retrieved the fully reduced spectrum from MAST.  The S/N per resolution element varies between $\sim 60$ at 1300 \AA\ to $\sim 20$ at 1700 \AA.  The STIS calibration pipeline yields spectra with a heliocentric wavelength scale.  We find that the STIS spectrum is slightly fainter than the FUSE spectrum in the region of overlap.  
{As we have determined that the FUSE spectrum is photometric,}
we scale the flux and error arrays of the STIS spectrum by 1.12.

\subsection{IUE Spectroscopy}\label{sec_IUE}

Instrumental details of the IUE satellite and its spectrograph can be found in \citet{Boggess:1978a, Boggess:1978b}.  Two observations of HD~177566 were obtained with the short-wavelength prime (SWP) camera in high-resolution mode through the large aperture, which yields a resolution of about 0.22 \AA\ \citep{Garhart:1997}.  We retrieved the fully-reduced spectra in ASCII format from the MAST website and combined them, weighting by exposure time.  We find that the fluxes of the IUE and rescaled STIS spectra, averaged over the wavelength range 1300--1600 \AA, are identical, confirming our decision to rescale the STIS data.

\subsection{UVES Spectroscopy}\label{sec_UVES}

HD~177566 was observed with the Ultraviolet and Visual Echelle Spectrograph \citep[UVES;][]{Dekker:2000} on the Very Large Telescope (VLT) at the European Southern Observatory (ESO).  Its fully calibrated spectrum was retrieved from the ESO Science Archive.  The observation was carried out using an $8\arcsec \times 0\farcs5$ slit with the blue and red arms centered on 390 and 580 nm, respectively.  While the UVES data-reduction pipeline corrects each echelle order for the blaze function and combines all orders into a single spectrum, the correction is imperfect, so we re-normalize the spectrum by fitting a low-order spline to the data.

The error bars in the UVES spectrum are overestimated by about a factor of two.  Fitting a horizontal line to a flat stretch of the continuum yields a reduced chi-squared value of $\chi^2_\nu \sim 0.2$, when we would expect a value of unity.  In our model fits, we identify a flat region of the continuum, measure the standard deviation of the flux and the mean value of the error bars in that region, and scale the latter to match the former.  With this correction, we find that the S/N per resolution element ranges from about 100 to 400 in the blue spectrum and between 400 and 500 in the red.

\section{Analysis}\label{sec:analysis}

\subsection{Interstellar Medium}\label{sec_ism}

The FUV spectrum of HD~177566 includes a variety of interstellar absorption features.  In the FUSE bandpass, molecular hydrogen is the dominant species, while high-ionization species are prominent at longer wavelengths.  Synthetic interstellar absorption spectra are computed using software written at the University of California, Berkeley, by M.\ Hurwitz and V.\ Saba. Given the column density, Doppler broadening parameter, and velocity of each component, the program computes a Voigt profile for each absorption feature and produces a high-resolution spectrum of optical depth versus wavelength.  Wavelengths, oscillator strengths, and other atomic data are taken from \citet{Morton:2003}.  For neutral hydrogen, we begin with the column density $\log N$(\hone) = $20.88 \pm 0.09$ derived by \citet{Diplas:Savage:1994} from high-resolution IUE data.  We find that the Lyman $\alpha$ line is better fit by a slightly lower column,  $\log N$(\hone) = 20.85, which is consistent with their result.  The high-order Lyman features are best reproduced by a Doppler parameter $b = 13$ \kms.  All of the ISM features are fit by eye, which is sufficient for our needs.  

According to the reddening maps of \citet{SFD:1998}, as recalibrated by \citet{Schlafly:Finkbeiner:2011}, the extinction in the direction of HD~177566 is \ebv\ = $0.084 \pm 0.002$.  We find that the shape of its FUSE spectrum is well reproduced by a \citet{Fitzpatrick:1999} extinction curve assuming \ebv\ = 0.04, extrapolated to the Lyman limit.  We will revisit the question of extinction below.

\subsection{Model Atmospheres}\label{sec_models}

We compute non-LTE stellar-atmosphere models using version 208 of the program TLUSTY \citep{Hubeny:Lanz:1995}.  We employ atomic models similar to those used by \citet{Lanz:Hubeny:2003} to compute their OSTAR grid.  Unlike \citeauthor{Lanz:Hubeny:2003}, we set the microturbulent velocity to zero in our model atmospheres.  
{
This change is motivated by several factors.  The star's metal lines are well fit by models that assume only instrumental line broadening.  We find no evidence of mass loss (\ie, metal lines with P~Cygni profiles), suggesting that the atmosphere is in radiative equilibrium.  Finally, the star is hot enough that its hydrogen and helium are mostly ionized, so there is no mechanism to drive convection in the atmosphere.  Overestimating the microturbulent velocity can have serious consequences:  \citet{Dixon:2019} found that models assuming a microturbulent velocity $\xi = 0$ \kms\ yield an Fe abundance for the post-AGB star Barnard 29 in M13 (for which \teff\ = 21,400 K) consistent with the cluster value, while models with $\xi = 10$ \kms\ yield an abundance 0.6 dex lower.
}

We begin by adopting the stellar parameters derived by \citet{Mello:2012}, \teff = $30{,}900 \pm 300$ K, $\log g = 3.83$, \abund{He} = $-0.91$, \abund{N} = $-4.77$, and \abund{O} = $-4.57$.  For other elements, we assume [M/H] = $-1.2$ and scaled-solar abundances.  Given a model atmosphere, we compute a synthetic spectrum using version 54 of the program SYNSPEC \citep{Hubeny:1988}.  For the FUSE data, synthetic spectra are convolved with a Gaussian of FWHM = 0.06 \AA\ to match the FUSE line-spread function.  For the STIS and IUE spectra, we use Gaussians with FWHM = 0.025 \AA\ and 0.22 \AA, respectively.  For the UVES spectrum, we model the spectrum in segments a few hundred \AA ngstroms wide; each segment is convolved with a Gaussian (whose FWHM ranges from 0.05 to 0.085 \AA) corresponding to a spectral resolution 
{$R \sim 65{,}000$ in the blue, $R \sim 75{,}000$ in the red.}

Before comparing our FUV models to the data, we scale them by the dust and ISM absorption models described above.  Finally, we scale the model to reproduce the continuum in a nearby (apparently) line-free region in the observed spectrum.  Because the optical spectra are normalized, we need not scale their synthetic spectra to account for reddening.

{
Our fitting routine is based on the function scipy.optimize.curve\_fit \citep[SciPy v1.16.0;][]{SciPy2020}.
Given a grid of synthetic spectra, prepared as described above, the program linearly interpolates among them---in one, two, or three dimensions, as appropriate---determining the best fit to the data via chi-squared minimization.  The uncertainties quoted for parameters derived from individual line fits are $1 \sigma$ errors computed from the covariance matrix returned by curve\_fit; we refer to these as statistical errors.
}

Continuum placement is the dominant uncertainty in our fits to the FUV spectra.  Weak absorption lines not included in our model may depress the apparent continuum.  Allowing our fitting routines to scale the model to the mean level of the ``pseudo-continuum'' thus underestimates the true continuum level.  
{
If the S/N ratio is high enough that one can assume that all of the apparent absorption lines are real spectral features (and if the iron abundance is low enough that the continuum is not depressed everywhere), then one can estimate the actual continuum level from the highest peaks in the observed spectrum, but this procedure is subjective and difficult to automate.  Instead, we perform each fit twice, once with the model continuum fixed as described above and again with the model scaled by a factor of 0.97 (3\% being a rough estimate of the difference between the actual and pseudo-continuum).
}
The difference in the two abundances is an estimate of the systematic error in our abundance estimates. We add this term and the statistical error in quadrature to compute our final error for a single absorption feature.  In most cases, the continuum uncertainty is the dominant contributor to the final error.  Because of its high S/N ratio and relative paucity of metal lines, the continuum is much better defined in the star's UVES spectrum, so we adopt the statistical error as the uncertainty in our optically-derived abundances.

\subsection{Stellar Parameters}\label{sec_params}

\begin{figure}
\epsscale{1.16}
\plottwo{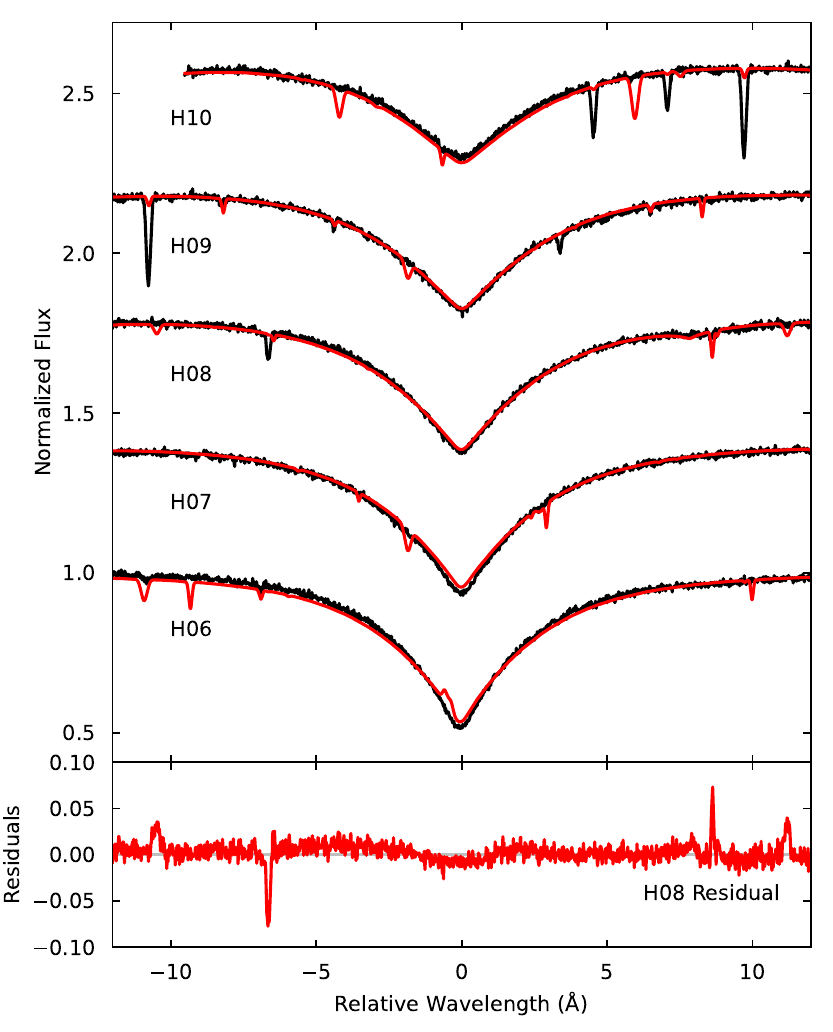}{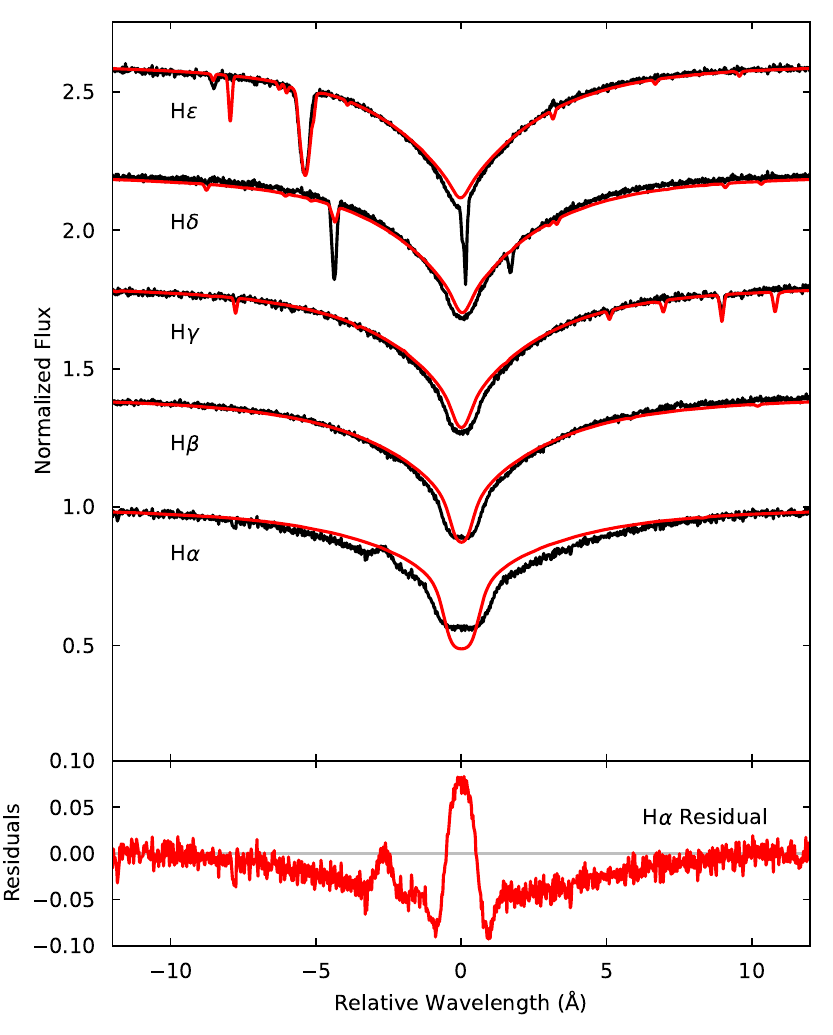}
\caption{Hydrogen lines in the optical spectrum of HD~177566.  Our best fit to the high-order features (left panel) is overplotted in red.}
\label{fig:balmer}
\end{figure}

We begin our analysis by fitting the high-order lines in the hydrogen Balmer series.  We generate a two-dimensional grid in effective temperature and surface gravity.  These features are well fit by a model with effective temperature \teff\ = 28,975 $\pm$ 23 K and surface gravity \logg\ = 3.68 $\pm$ 0.002.  Unfortunately, as shown in \figref{fig:balmer}, neither this model grid nor any other that we have employed can reproduce the star's low-order Balmer lines, which are broader than our models and have flatter cores.  We assume that these lines are shaped by processes in regions of the star's atmosphere beyond those modeled by TLUSTY.

\begin{figure}
\epsscale{1.16}
\plottwo{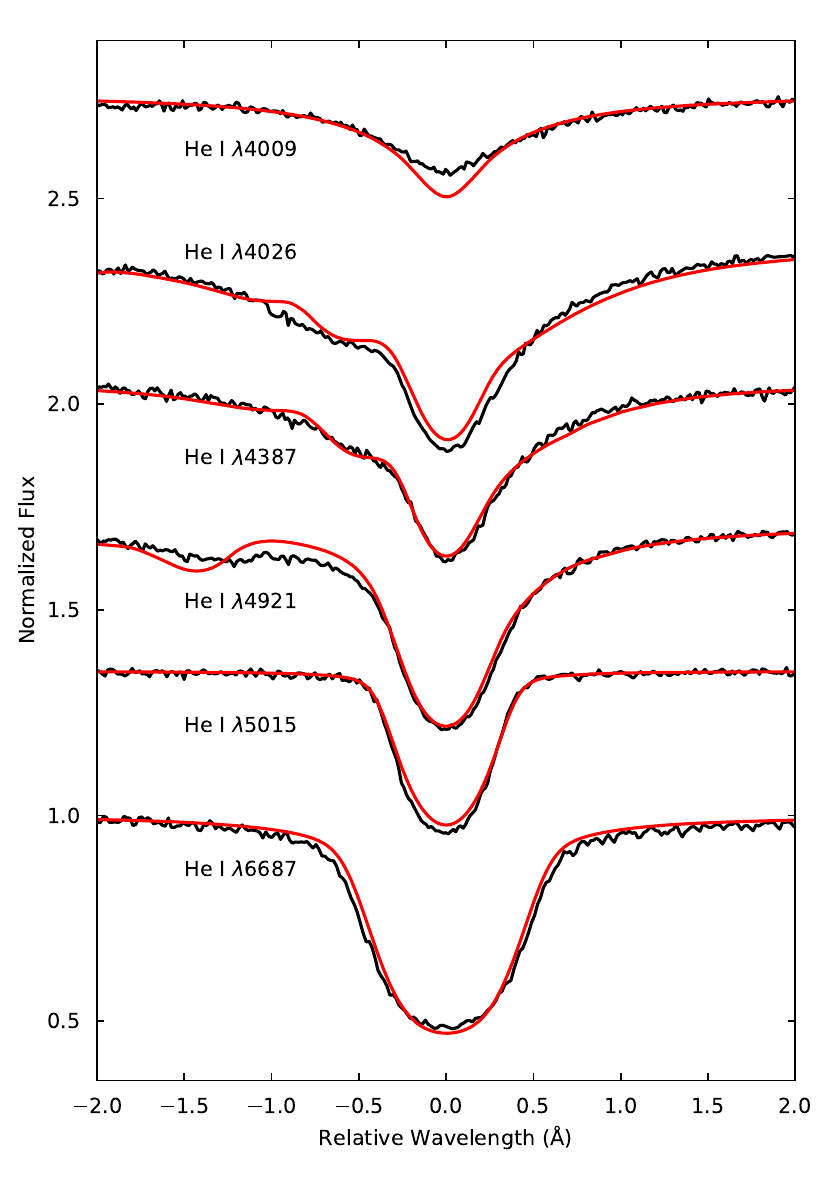}{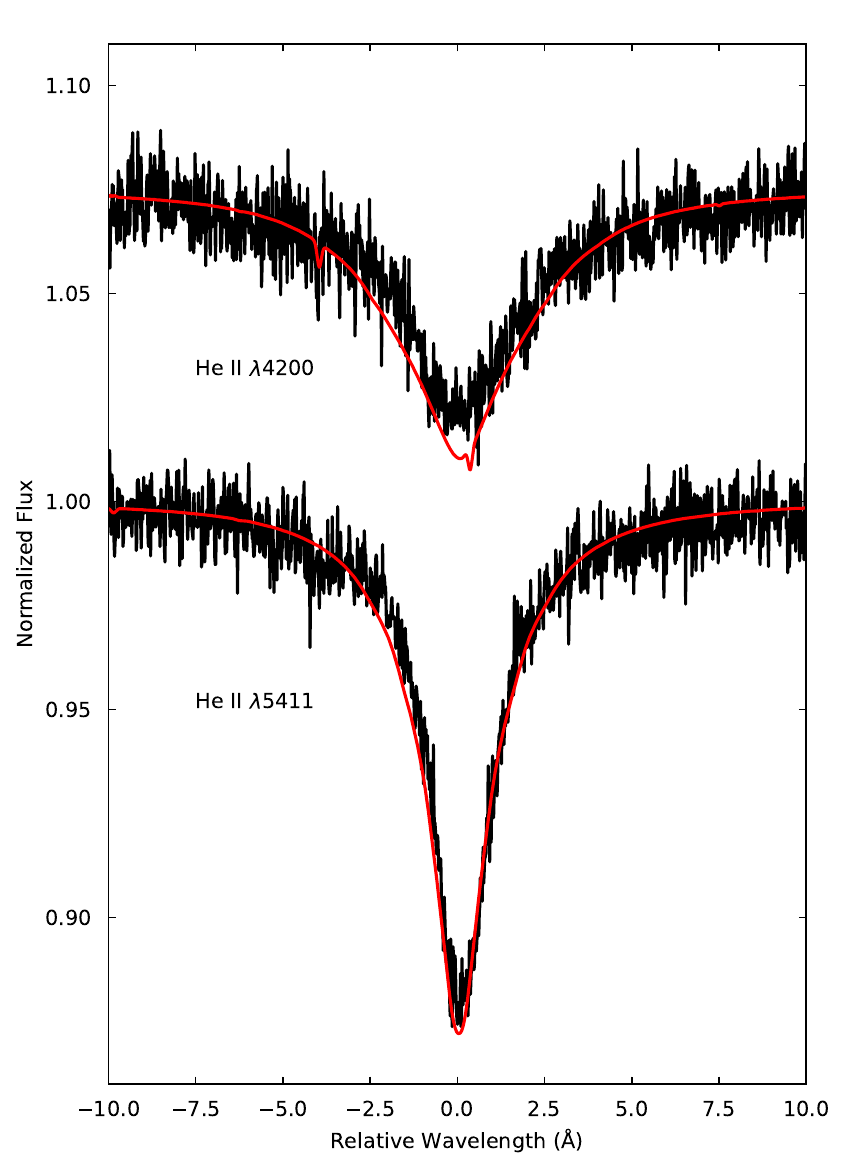}
\caption{Helium lines in the optical spectrum of HD~177566.  Our best-fit synthetic spectrum is overplotted in red.}
\label{fig:hetwo}
\end{figure}

For a second estimate of the stellar parameters, we follow \citet{Mello:2012}, simultaneously fitting the \heone\ lines at 4009, 4026, 4387, 4921, 5015, and 6678 \AA\ and the \hetwo\ lines at 4200 and 5411 \AA.  (The \hetwo\ $\lambda 4541$ line falls in the gap between the blue and red spectra.)  These lines were selected to avoid the wind contamination seen in several hot stars in the Mello et al.\ sample.  Given a three-dimensional grid in effective temperature, surface gravity, and helium abundance, our fitting routine returns best-fit values of \teff\ = $32{,}887 \pm 22$ K, $\log g = 3.79 \pm 0.004$, and \abund{He} = $-0.86 \pm 0.002$.  The data and best-fit model are presented in \figref{fig:hetwo}.

\begin{figure}
\plotone{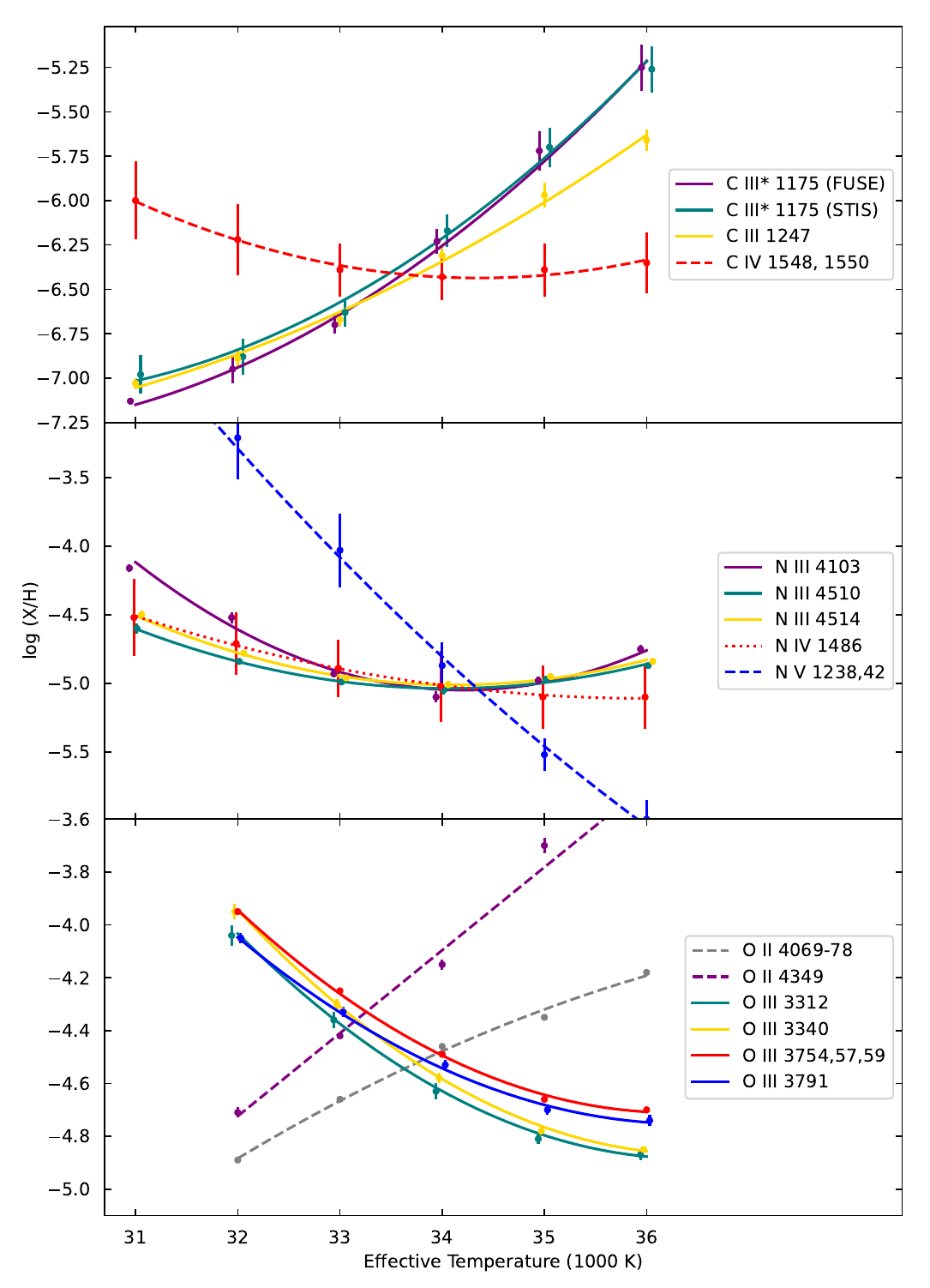}
\caption{Deriving the effective temperature from the CNO lines.  Points with error bars represent the abundance derived from model fits to each absorption feature.  Curves are low-order polynomial fits to the measured points.}
\label{fig:teff}
\end{figure}

For a third estimate, we follow \citet{Dixon:M5:2024, Dixon:2025}, using the absorption features of multiple ionization states of CNO to derive the star's effective temperature.  Consider \figref{fig:teff}.  We generate a set of model atmospheres with \teff\ = 31,000 K and \abund{C} between $-7.5$ and $-5.0$.  We begin by fitting the \cthree\ $\lambda 1247$ feature to determine the carbon abundance.  We repeat using models with temperatures increasing in steps of 1000 K to 36,000 K.  The resulting carbon abundances are plotted as red points and connected by a low-order polynomial.  Vertical bars represent the uncertainties returned by the fitting routine.  As the temperature rises, the fraction of carbon in the form of \cthree\ falls, requiring a higher carbon abundance to reproduce the observed feature.  We repeat this procedure for the \cthree* $\lambda 1175$ multiplet (seen by both FUSE and STIS) and the \cfour\ $\lambda 1550$ doublet.  We employ the same process for selected N and O lines (middle and bottom panels of \figref{fig:teff}).  We see that the curves cross at a temperature near 34,000 K.  

Because the star has few metal lines in the optical, both \citet{Kendall:1994} and \citet{Mello:2012} used its hydrogen lines to determine its effective temperature.  Our fits to the star's hydrogen lines yield results similar to theirs.  It was only when we began measuring the CNO abundances that we discovered the need for a higher effective temperature.  We thus adopt the stellar parameters from our fits to the helium lines:  \teff = $33{,}000 \pm 1000$ K, $\log g = 3.79 \pm 0.11$, and \abund{He} = $-0.86 \pm 0.05$.  Error bars for \teff\ represent the difference between the helium and CNO values, those for \logg\ the difference between the helium and hydrogen values, and those for \abund{He} the difference between our helium abundance and that of \citeauthor{Mello:2012}  These and other stellar parameters are presented in Table \ref{tab:stellar_params}.

\begin{deluxetable}{lcccc}[t]
\tablecaption{Stellar Parameters \label{tab:stellar_params}}
\tablehead{
\colhead{Parameter} & \colhead{Value}
}
\startdata
\teff\ (K) & $33{,}000  \pm 1000$ \\
\logg\ [cm s$^{-2}$] & $3.79 \pm 0.11$ \\
\abund{He} & $-0.86 \pm 0.05$ \\
$R_*/R_{\sun}$ & $1.72 \pm 0.13$ \\
$M_*/M_{\sun}$ & $0.67 \pm 0.20$ \\
$\log (L_*/L_{\sun})$ & $3.50 \pm 0.08$ \\
\enddata
\end{deluxetable}

\subsection{Metal Abundances}\label{sec_abundance}

Thirteen additional elements, C, N, O, Ne, Mg, Al, Si, P, S, Cl, Fe, Ni, and Ge, exhibit absorption lines in the spectrum of HD~177566.  We derive their abundances by fitting synthetic spectra to the features listed in Table~\ref{tab:lines_cno} (for CNO) and Table~\ref{tab:lines_other} (everything else), found in the Appendix.  For each element, we generate a set of stellar-atmosphere models with a single temperature but a range of abundance values.  The final abundance and its uncertainly represent the weighted mean and weighted standard deviation of our individual measurements (which has the practical effect of favoring our optical results).  Results are presented in Table \ref{tab:abundance} and plotted in \figref{fig_abundance}.

{
The dependence of our CNO abundances on the assumed effective temperature is illustrated in \figref{fig:teff}.  Though our value of \teff\ is 2100 K higher than \cite{Mello:2012}, our He, N, O, and Si abundances are consistent with theirs, while our Mg abundance is lower by 0.4 dex.  Models with \logg\ = 4.0 yield CNO abundances consistent with those presented in Table \ref{tab:abundance}.
}

Notes on individual species follow.

\begin{figure}
\plotone{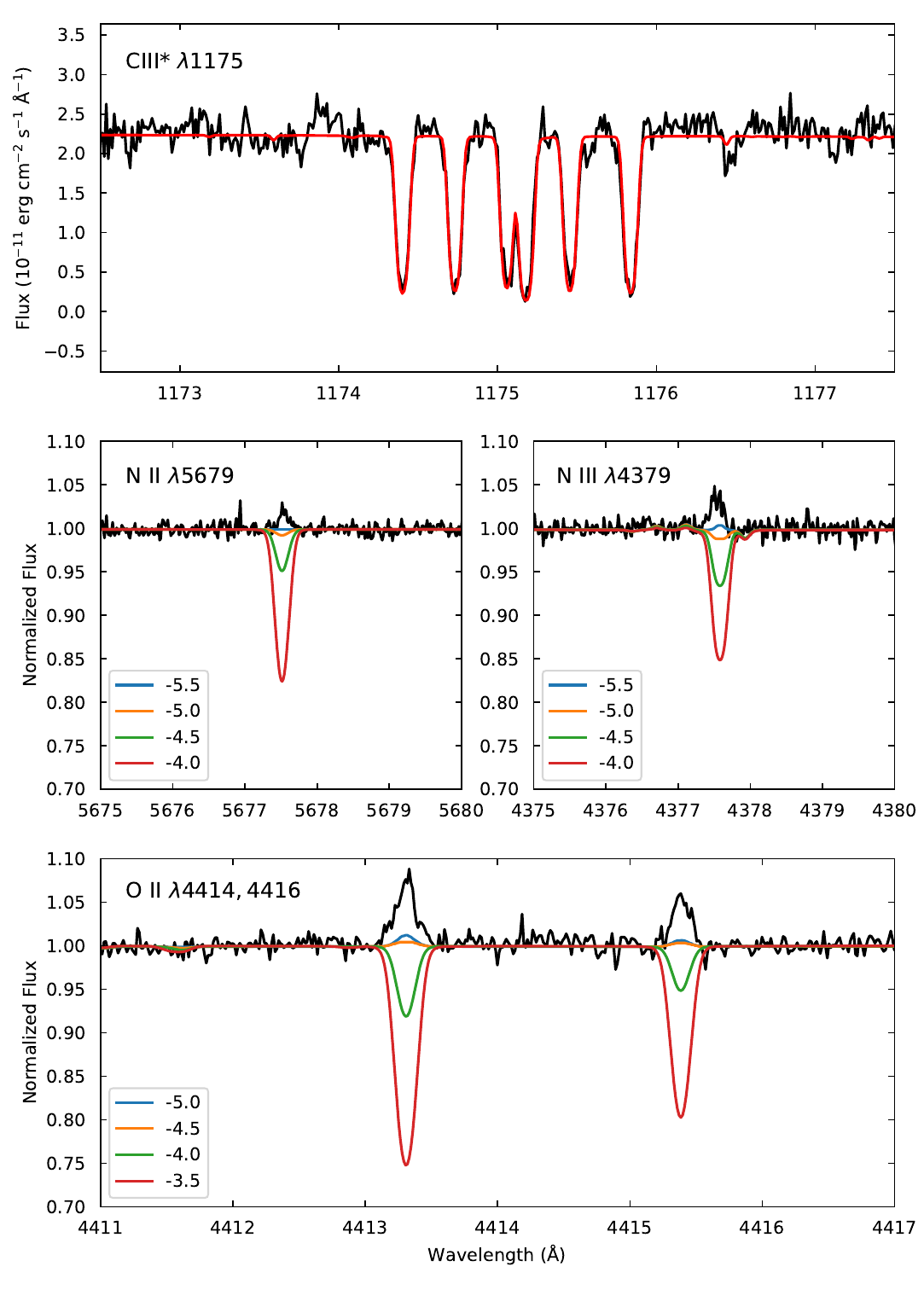}
\caption{CNO features in the spectrum of HD~177566.  Top panel: \cthree * $\lambda 1175$ multiplet in the STIS spectrum.  Though the carbon abundance is low, these lines are strong and well reproduced by our best-fit model (red curve).  Middle panel: the \ntwo\ $\lambda 5679$ (left) and \nthree\ $\lambda 4379$ emission features are not reproduced by our models.  Bottom panel: for the \otwo\ $\lambda \lambda 4414, 4416$ doublet, our models yield emission features only for the lowest abundances, and these features are too weak.  Colored curves represent synthetic spectra derived using BSTAR model atoms with \abund{X} as labeled. }
\label{fig_emission}
\end{figure}

{\em Carbon:}\/
{
While the star's carbon abundance is quite low, the value seems secure.  A plot of our best fit to the \cthree* $\lambda 1175$ multiplet in the STIS spectrum is presented in the top panel of \figref{fig_emission}.  The line profiles are well reproduced by our synthetic spectrum; no additional line broadening is required.
}

{\em Nitrogen:}\/  
{
The \ntwo\ $\lambda 5679$ and \nthree\ $\lambda 4379$ lines are weak emission features (middle panel of \figref{fig_emission}).  Our initial set of synthetic spectra, computed using the OSTAR atomic models of \citet{Lanz:Hubeny:2003}, predict these features to be in absorption.  Emission features arise when ions in excited states cascade to lower energy levels; they can be reproduced only by atomic models that include the relevant high-energy states.  The OSTAR model for \ntwo\ contains 18 energy levels and considers 97 allowed transitions.  In an attempt to reproduce the \ntwo\ $\lambda 5679$ feature, we generate a second set of model atmospheres using the BSTAR atomic models of \citet{Lanz:Hubeny:2007}, which are more sophisticated.  The BSTAR model for \ntwo\ contains 32 energy levels and considers 278 allowed transitions.  Unfortunately, even this model cannot reproduce the \ntwo\ $\lambda 5679$ emission feature, which we exclude from our analysis.
}

{
Both the OSTAR and the BSTAR models employ a \nthree\ model atom with 25 energy levels and 184 allowed transitions, but a more complete model, with 40 energy levels and 403 allowed transitions, is available on the \href{https://tlusty.oca.eu/tlusty}{TLUSTY website}.\footnote{Note that the file n3\_40+9lev.dat contains an error.  In the final six lines specifying bound-free transitions, the IFANCY parameter (column 4) is set to 2 but should be 0.}  Again, this model is unable to reproduce the \nthree\ $\lambda 4379$ emission feature, which we exclude from our analysis.
}

{\em Oxygen:}\/  
{
The file gfVIS99.dat, which is used by SYNSPEC and is available from the \href{https://tlusty.oca.eu/tlusty/Synspec49/synspec-line.html}{Synspec Line Lists Repository}, contains two entries for the \otwo\ $\lambda 4075$ feature.  They have nearly identical wavelengths (4075.862 \AA\ and 4075.866 \AA) and atomic parameters.  Only the 4075.862 \AA\ feature appears in the NIST Atomic Spectra Database \citep{Kramida:2024}.  We delete the 4075.866 \AA\ feature from the line list before generating our synthetic spectra.
}

{
The \otwo\ lines at 3945, 3954, 3973, 3982, 4414, 4416, 4941, and 4416 \AA\ are seen in emission.  Our initial set of synthetic spectra predict these features to be in absorption.  The OSTAR model for \otwo\ contains 21 energy levels and considers 125 allowed transitions; the corresponding BSTAR model contains 36 energy levels and considers 346 allowed transitions. For most values of the oxygen abundance, spectra generated from model atmospheres using BSTAR \otwo\ models predict these features to be in absorption.  As illustrated in the bottom panel of \figref{fig_emission}, models with \abund{O} = $-5.0$ and (for some lines) $-4.5$ do exhibit  \otwo\ emission, but the lines are much weaker than observed.  We exclude these features from our analysis.
}

{\em Neon:}\/ 
{
The \netwo\ features at 3334, 3344, 3355, 3664, and 3694 \AA\ are seen in absorption, but our initial models predict emission in the 3334, 3364, and 3694 \AA\ lines.  Synthetic spectra generated from model atmospheres using the BSTAR \netwo\ model atom yield synthetic spectra with all five lines in absorption.  We use these models and all five lines to derive the neon abundance.
}

{\em Chlorine, Germanium:}\/ We lack model atoms for these elements, so treat them in LTE.

\begin{deluxetable}{lcccc}
\caption{Photospheric Abundances \label{tab:abundance}}
\tablehead{
\colhead{Species} & \colhead{HD~177566} & \colhead{Features\tablenotemark{a}} & \colhead{M13\tablenotemark{b}} & \colhead{Sun\tablenotemark{c}}
}
\startdata
He & $-0.90 \pm 0.10$ & \phn8 & \nodata & $-1.09 \pm 0.01$ \\
C  & $-6.65 \pm 0.06$ & \phn4 & $-5.55 \pm 0.14$ & $-3.54 \pm 0.04$ \\
N  & $-4.97 \pm 0.09$ & 13 & $-4.89 \pm 0.13$ & $-4.17 \pm 0.07$ \\
O  & $-4.43 \pm 0.15$ & 15 & $-4.28 \pm 0.13$ & $-3.31 \pm 0.04$ \\
Ne & $-5.62 \pm 0.22$ & \phn5 & \nodata & $-3.94 \pm 0.05$ \\
Mg & $-6.05 \pm 0.06$ & \phn1 & $-5.76 \pm 0.06$ & $-4.45 \pm 0.03$ \\
Al & $-7.47 \pm 0.06$ & \phn2 & $-7.09 \pm 0.13$ & $-5.57 \pm 0.03$ \\
Si & $-5.78 \pm 0.12$ & 10 & $-5.61 \pm 0.08$ & $-4.49 \pm 0.03$ \\
P  & $-8.29 \pm 0.10$ & \phn3 & \nodata & $-6.59 \pm 0.03$ \\
S  & $-6.20 \pm 0.24$ & 15 & \nodata & $-4.88 \pm 0.03$ \\
Cl & $-9.18 \pm 0.11$ & \phn2 & \nodata & $-6.69 \pm 0.20$ \\
Fe & $-6.18 \pm 0.17$ & \phn6 & $-6.05 \pm 0.07$ & $-4.54 \pm 0.04$ \\
Ni & $-7.05 \pm 0.16$ & \phn4 & $-7.35 \pm 0.09$ & $-5.80 \pm 0.04$ \\
Ge & $-10.81 \pm 0.08$ & \phn2 & \nodata & $-8.38 \pm 0.10$ \\
\enddata
\tablecomments{Abundances relative to hydrogen: \abund{X}.}
\tablenotetext{a}{Number of features fit for each element.  Some are blends or multiplets; see Appendix for details.} 
\tablenotetext{b}{M13 values from \citet{Meszaros:2015}, except for nickel, which is from \citet{Cohen:Melendez:2005}.}
\tablenotetext{c}{Solar values from \citet{Asplund:2021}.}
\end{deluxetable}

\section{Discussion} \label{sec:discussion}

\subsection{Diffuse Lines of Helium}\label{sec:anomalies}

\begin{figure}
\epsscale{1.16}
\plottwo{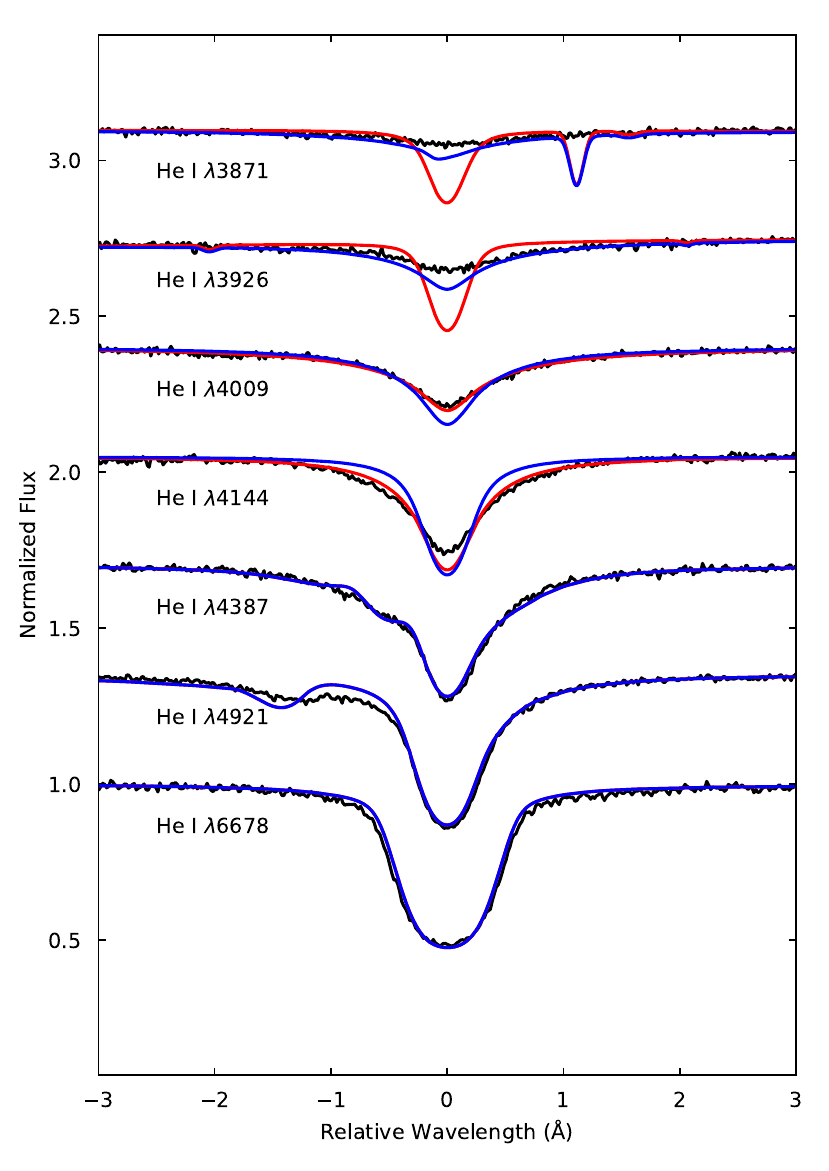}{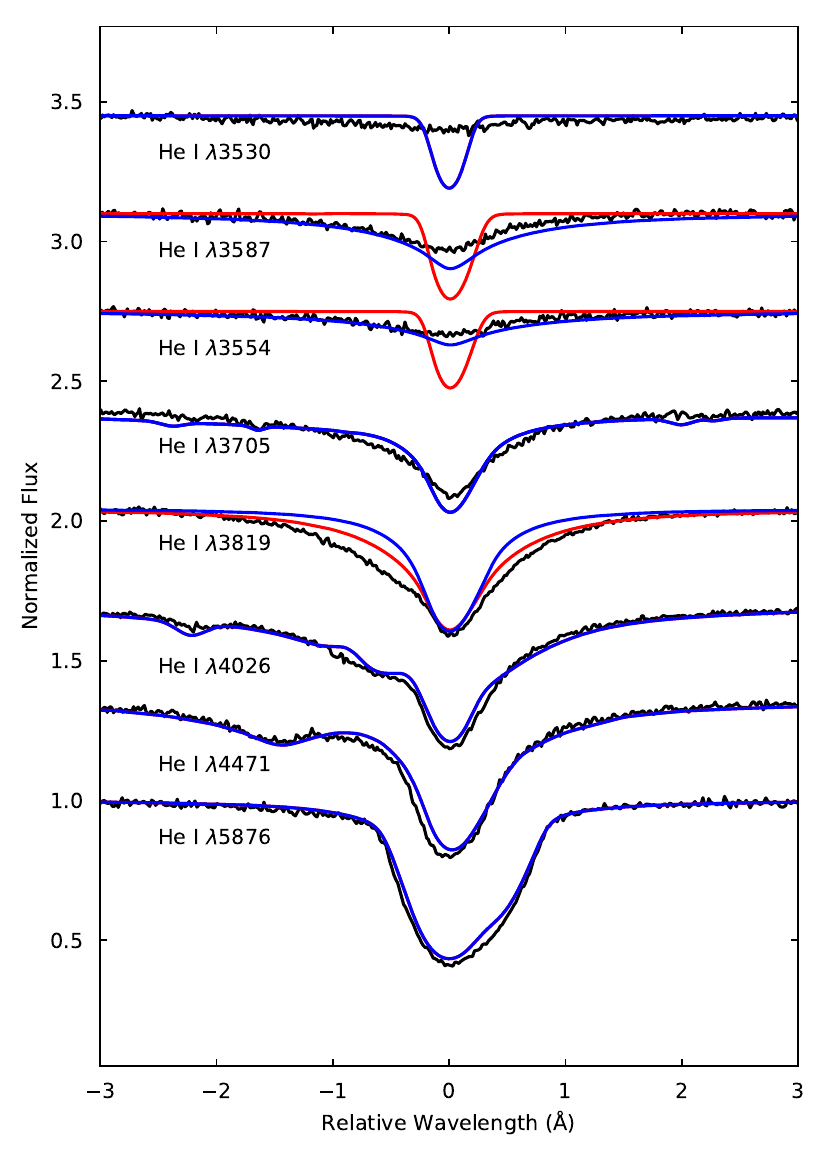}
\caption{Diffuse helium lines in the optical spectrum of HD~177566.  Singlet lines are on the left, triplet lines on the right.  The observed spectrum is plotted in black.  Synthetic spectra generated with version 54 of SYNSPEC are plotted in red.  In blue are synthetic spectra generated using a new version of SYNSPEC, modified as described in the text.}
\label{fig:diffuse}
\end{figure}

The diffuse lines of \heone\ require special treatment.  These features are caused by electron transitions from the lowest P orbital to an excited D state.  Within SYNSPEC, the line profiles of \citet{Shamey:1969} and \citet{Barnard:1974} are used for the lines at 4026, 4387, 4471, and 4921 \AA.  For 20 other lines, Stark broadening parameters are taken from the compilation of \citet{Dimitrijevic:1984} or computed using the formula of \citet{Freudenstein:1978}.  To improve these line fits, we have replaced the \citeauthor{Dimitrijevic:1984} and \citeauthor{Freudenstein:1978} line-broadening parameters with values from the Stark-B database of \citet{stark-b}, which includes data from \citet{Dimitrijevic:1989, Dimitrijevic:1990}. For 31 features, we use the tabulated values for an electron density $\log N(e^{-}) = 16$.  For ten additional features, we extrapolate from the tabulated values for $\log N(e^{-}) = 14$ and 15 (and multiply the result by three).  As shown in \figref{fig:diffuse}, the new line profiles generally yield improved fits to the data.  Exceptions are the features at 3819, 4009, and 4144 \AA, for which the original line-broadening parameters are slightly larger than the new ones, and \heone\ $\lambda 3530$, which is modeled with a Voigt profile in both versions of the code.

\subsection{Photospheric Abundances}

\begin{figure}
\epsscale{1.18}
\plotone{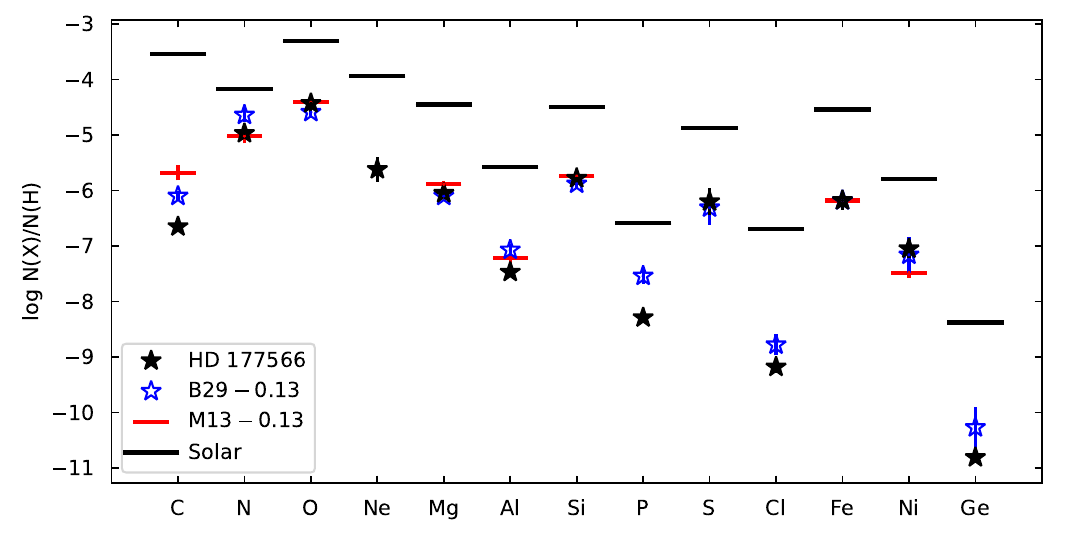}
\caption{Photospheric abundances of HD~177566 (black stars), Barnard~29 (blue stars), first-generation RGB stars in M13 (red lines), and the solar photosphere (black lines).  Barnard~29 and cluster values are scaled to match the iron abundance of HD~177566.  M13 values are from \citet{Meszaros:2015}, except for nickel, which is from \citet{Cohen:Melendez:2005}.  Barnard~29 values from \citet{Dixon:2019}.  Solar values from \citet{Asplund:2021}.}
\label{fig_abundance}
\end{figure}

Table \ref{tab:abundance} and \figref{fig_abundance} compare the photospheric abundances of HD~177566 with a population of Galactic halo stars and with the sun.  For the halo stars, we use the first-generation RGB stars in the globular cluster M13 \citep[NGC~6205;][]{Meszaros:2015}, which has an iron abundance similar to that of HD~177566.  We also plot the abundances of the post-AGB star Barnard~29 \citep{Dixon:2019}, which is a member of M13.  In \figref{fig_abundance}, we subtract 0.13 dex from both the cluster and Barnard~29 abundances to match the iron abundance of HD~177566.  The HD~177566 and (adjusted) M13 values agree within the errors for all elements except carbon, which is underabundant by about 1 dex.  

Except for C and N, the abundances of Barnard~29 also match the cluster values, suggesting that its chemistry has changed little since its departure from the RGB---and thus that its P, S, Cl, and Ge abundances are representative of the cluster values, which have not yet been measured.  We see that, relative to the (adjusted) values of Barnard~29, HD~177566 is also underabundant in P, Cl, and Ge.

\citet{Meszaros:2015} find a clear correlation in the abundance of carbon with effective temperature in M13, which they
interpret as a sign of ``deep mixing,'' a nonconvective mixing process that results in a steady depletion of the surface carbon abundance and an enhancement in the nitrogen abundance in low-mass stars as they evolve up the
RGB \citep{Gratton:2004}. The abundance pattern seen in Barnard~29---C depletion and N enhancement relative to the
average value for first-generation stars on the RGB---suggests that the star experienced the full impact of deep mixing.  The photosphere of HD~177566 is not enhanced in nitrogen relative to the cluster, but its carbon abundance is even lower than that of Barnard~29, which suggests that some other process may have been at work in this star.  

\subsection{Stellar Mass and Luminosity}\label{sec:luminosity}

We can derive the star's radius, and from this its mass and luminosity, by comparing its observed and predicted flux. 
The spectral irradiance of HD~177566 has been measured in a range of optical and NIR bands:  
$V = 10.20 \pm 0.024$, $B - V = -0.20 \pm 0.015$, $U - B = -1.00 \pm 0.019$ \citep{Hill:1970};
$ G = 10.140 \pm 0.003$ from Gaia DR3 \citep{Gaia_DR3:2023};
and $J = 10.588 \pm 0.023$, $H = 10.701 \pm 0.025$, $K = 10.746 \pm 0.021$ from 2MASS \citep{Skrutskie:2006}.

\begin{figure}
\epsscale{1.18}
\plotone{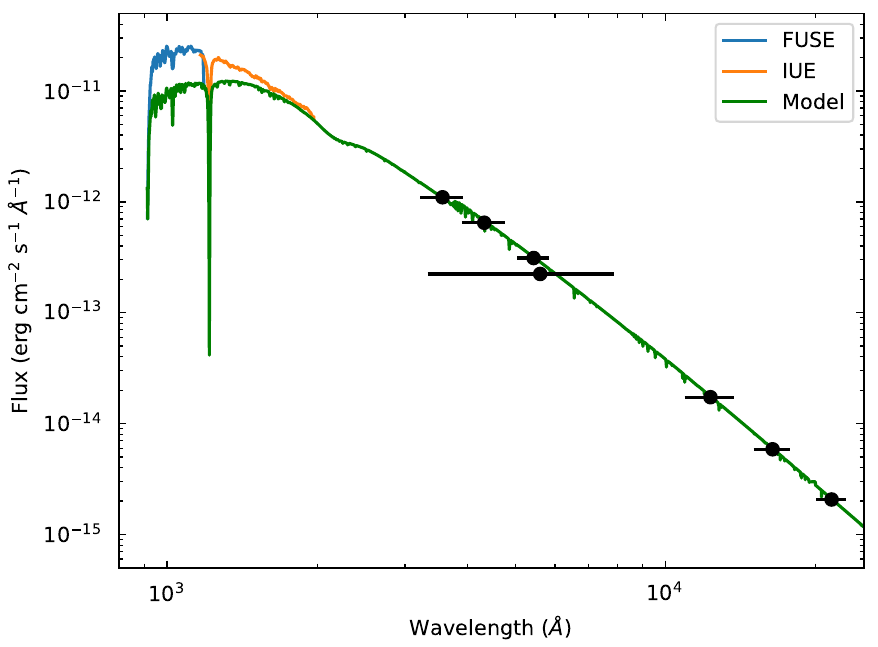}
\caption{Spectral-energy distribution of HD~177566.  The blue and orange curves are the FUSE and IUE spectra, respectively.  Black crosses are optical and NIR magnitudes from the literature, expressed in units of flux.  The green curve is our best-fit model.  Scaled to reproduce the optical and NIR data, it underpredicts the flux at FUV wavelengths by a factor of two.  Both the observed and model spectra have been smoothed by 10 \AA\ for this figure.}
\label{fig:mags}
\end{figure}

To model the stellar continuum, we use our best-fit NLTE model with the abundances derived in Section \ref{sec_abundance}.  Synthetic stellar magnitudes are computed using the Python packages synphot and stsynphot \citep{synphot:2018, stsynphot:2020}.  For the $U$, $B$, and $V$ filters, we use the latest Vega spectrum from CALSPEC \citep[\texttt{alpha\_lyr\_stis\_011.fits};][]{Bohlin:2014}.  For Gaia, we follow the recipe of \citet{Riello:2021}, using the \texttt{alpha\_lyr\_mod\_002.fits} spectrum of Vega scaled to $3.62286 \times 10^{-11}$ W m$^{-2}$ nm$^{-1}$ at 5500 \AA.  For the 2MASS filters, we follow \citet{Cohen:1992}, adopting the \citet{Kurucz:1991} spectrum of Vega scaled to $3.44 \times 10^{-9}$ \flux\ at 5556 \AA.  For this exercise, we adopt the dust models of \citet{Gordon:2023}\footnote{These models incorporate the work of \citet{Cardelli:1989, Fitzpatrick:2019, Gordon:2021}; and \citet{Decleir:2022}.} as implemented in the Python package dust\_extinction \citep{Gordon:2024}.  We assume $R_V = 3.1$, a value appropriate for low-density Galactic sight lines. Free parameters in the fit are the scale factor of the model continuum and the extinction parameter \ebv.  The best-fit model, with a scale factor $\phi = (1.01 \pm 0.01) \times 10^{-20}$ and an extinction \ebv = $0.095 \pm 0.005$, is plotted in \figref{fig:mags}.  Our extinction is similar to the value \ebv\ = $0.084 \pm 0.002$ reported by \citet{Schlafly:Finkbeiner:2011} but more than twice that derived from the shape of the FUSE continuum (Section \ref{sec_ism}).

In \figref{fig:mags}, the green curve represents the model, and black crosses indicate the optical and NIR magnitudes.  Though the model is scaled to fit only the optical and NIR data, we extend it into the FUV for comparison to the FUSE and IUE spectra, which are plotted in blue and orange, respectively.  We see that it underpredicts the flux at FUV wavelengths by roughly a factor of two.  This discrepancy cannot be reduced by raising the stellar temperature; to match the observed optical and NIR magnitudes, hotter models require more extinction, which depresses the FUV flux.  An extinction curve  with $R_v \sim 4.8$, corresponding to a line of sight with fewer small dust grains than average, reproduces the level, if not the shape, of the star's FUV spectrum, but there is no evidence that the line of sight towards HD~177566 is unusual.
{
Our dust model may be incomplete; \citet{Gordon:2024} point out that standard extinction curves describe only the average behavior of the dust absorption in the ISM, and individual sight lines may vary.
}

In the synthetic spectra generated by SYNSPEC, the flux is expressed in terms of the flux moment, $H_\lambda$.  If the star's radius and distance are known, then the scale factor required to convert the model spectrum to the flux at earth is $\phi = 4 \pi (R_*/d)^2$ \citep{Kurucz:1979}.  \citet{Anders:2022} derived Bayesian stellar parameters, distances, and extinctions for 362 million stars from Gaia EDR3 data cross-matched with the photometric catalogues of Pan-STARRS1, SkyMapper, 2MASS, and AllWISE.  Their distance to HD~177566 is $d = 1371.8^{+127.0}_{-78.5}$ pc.  Adopting this distance and our scale factor, we derive a stellar radius $R_*/R_{\sun} = 1.72 \pm 0.13$.  Applying our adopted surface gravity ($\log g = 3.79 \pm 0.11$), we find that the stellar mass is $M_* / M_{\sun} = 0.67 \pm 0.20$.  Finally, combining the stellar radius with our best-fit effective temperature (\teff\ = 33,000 $\pm$ 1000 K), we derive a stellar luminosity $\log L / L_{\sun} = 3.50 \pm 0.08$.  (The uncertainties on these values are calculated via simple propagation of errors.)

Our value for the luminosity of HD~177566 is considerably greater than the value $\log (L/L_{\sun}) = 2.59 \pm 0.14$ derived by \citet{Partha:2020}.  This discrepancy is apparently due to an arithmetic error.  Using the numbers in Tables 1 and 2 of their paper, together with the bolometric correction and equations of \citet{Cox:2000}, we find $M_V = -0.926$ and $\log L / L_{\sun} = 3.55$.

\subsection{Evolutionary Status}

\begin{figure}
\epsscale{1.18}
\plotone{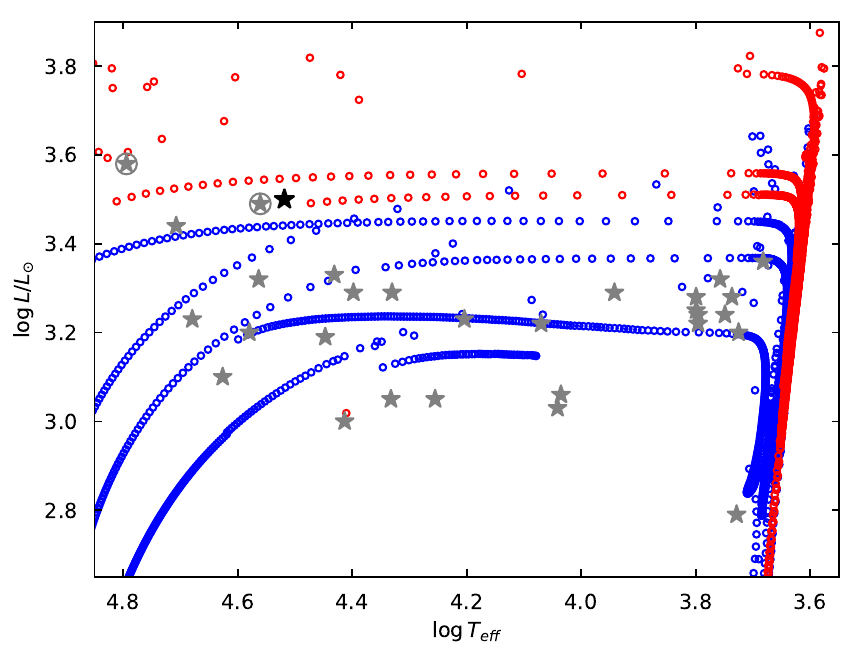}
\caption{Theoretical Hertzsprung--Russell diagram.  HD~177566 is plotted in black; additional stars from the literature are plotted in gray.  Overplotted are post-horizontal branch (HB) evolutionary tracks for stars with [Fe/H] = $-1.5$ from \citet{Moehler:2019}.  Tracks for stars evolving from the blue HB are represented by blue dots, those evolving from the red HB by red dots.  Each dot represents a time step of 1 Myr.}
\label{fig:cmd}
\end{figure}

Using the values in Table \ref{tab:abundance}, we derive a metallicity of [Fe/H] = $-1.64$ for HD~177566.   In \figref{fig:cmd}, we plot post-AGB evolutionary tracks for globular-cluster stars with [Fe/H] = $-1.5$ from \cite{Moehler:2019}.  Blue points represent the tracks for stars that evolved from the blue HB (with ZAHB masses between 0.53 and 0.65 \msun).  Red points represent the tracks for more massive stars that evolved from the red HB.  HD~177566 is represented as a black star, and additional cluster post-AGB stars from the literature are plotted in gray \citep{Moehler:2019, Ciardullo:2022, Dixon:3DU:2024}.  The luminosity and temperature of HD~177566 suggest that it is a post-AGB star that evolved from an object 
{
on the low-mass end of the red HB.  Such a star is predicted to have a final mass near 0.540 \msun; the measured mass of HD~177566, $0.67 \pm 0.20$ \msun, is consistent with this prediction.
}

The two circled stars in \figref{fig:cmd} are GJJC-1 in NGC 6656 \citep[the hotter of the two;][]{Gillett:1989} and K648 in NGC 7978 \citep{Otsuka:2015}, both of which are the central stars of planetary nebulae (CSPN).  Of the two, only K648 has been studied in detail.  The proximity of HD~177566 to those CSPN suggests that it might also have a PN, but this unlikely.  The star's carbon abundance, $\log N({\rm C})/N({\rm O}) = -2.18 \pm 0.21$, is three orders of magnitude lower than that of K648.  In fact, it is lower than that of any globular-cluster post-AGB star in the sample of \citet{Dixon:3DU:2024}, suggesting that the star did not ascend the AGB far enough to undergo significant third dredge-up, let alone to eject a PN.

\section{Conclusions} \label{sec:conclusion}

We have analyzed archival FUV and optical spectra of the hot halo star HD~177566.  Our estimates of the star's effective temperature, surface gravity, and helium abundance are similar to those of previous observers.  We derive the abundance of 13 chemical elements; these values are consistent with those of other halo stars, save for carbon, which is underabundant by about 1 dex.  The low-order lines of the star's hydrogen Balmer series are not well reproduced by our models, as they are broader and have flatter cores than our models.  The diffuse lines of \heone\ are often broader than predicted; the use of more recent line-broadening parameters significantly improves the fit.  Comparing our best-fit model to the star's optical and near-IR magnitudes yields an extinction \ebv = $0.095 \pm 0.005$, consistent with literature values, but the resulting model underpredicts the star's FUV flux by a factor of two.  The star's effective temperature and luminosity place it on the post-AGB evolutionary tracks of stars that evolved from the red horizontal branch, but its low C/O ratio indicates that it did not experience significant third dredge-up while on the AGB.

\begin{acknowledgments}

%
This work has made use of
NASA's Astrophysics Data System (ADS); 
the SIMBAD database, operated at CDS, Strasbourg, France; 
the Mikulski Archive for Space Telescopes (MAST), hosted at the Space Telescope Science Institute, which is operated by the Association of Universities for Research in Astronomy, Inc., under NASA contract NAS5-26555; 
data obtained from the European Southern Observatory (ESO) Science Archive Facility with DOI \url{https://doi.org/10.18727/archive/50}; and
data from the European Space Agency (ESA) mission
Gaia (\url{https://www.cosmos.esa.int/gaia}), processed by the Gaia
Data Processing and Analysis Consortium (DPAC;
\url{https://www.cosmos.esa.int/web/gaia/dpac/consortium}).
Funding for the DPAC has been provided by national institutions,
in particular the institutions participating in the Gaia Multilateral Agreement.
This work was performed, in part, during research leave granted by the STScI Science Mission Office, whose support is gratefully acknowledged.
Its publication is supported by the STScI Director's Discretionary Research Fund.

\end{acknowledgments}





%

\facilities{ESO(UVES), FUSE, HST(STIS), IUE}

\software{
dust\_extinction \citep{Gordon:2024},
stsynphot \citep{stsynphot:2020},
synphot \citep{synphot:2018}
}

\clearpage

\appendix

\section{Selected Features in the Spectrum of HD~177566}

\startlongtable
\begin{deluxetable*}{lcDrcc}
\tablecaption{Selected CNO Features \label{tab:lines_cno}}
\tablehead{
\colhead{Ion} & \colhead{$\lambda_{\rm lab}$} & \multicolumn2c{$\log gf$} & \colhead{$E_l$} & \colhead{Abundance} & \colhead{Instrument} \\
\colhead{} & \colhead{(\AA)} & \multicolumn2c{} & \colhead{(cm$^{-1}$)}
}
\decimals
\startdata
\cthree  & 1175.665\tablenotemark{a} & 0.39 & 52,419.400 & $-6.69 \pm 0.05$ & FUSE \\
              & 1175.665\tablenotemark{a} & 0.39 & 52,419.400 & $-6.61 \pm 0.08$ & STIS \\
\cthree   &  1247.383  &  -0.31  &  102,352.040  & $-6.65 \pm 0.04$ & STIS \\
\cfour &  1548.195  &  -0.42  &  0.000  & $-6.38 \pm 0.15$ & STIS \\
          &  1550.772  &  -0.72  &  0.000  & \nodata  &  \\
\nthree  &   979.905  &  -0.10  &  101023.904  & $ -4.87 \pm 0.06 $ & FUSE \\
         &  1005.993  &  -0.81  &  131004.300  & $ -5.11 \pm 0.05 $ &  \\
         &  1006.036  &  -1.12  &  131004.300  & \nodata &  \\
         &  1182.971  &  -0.92  &  145875.697  & $ -4.84 \pm 0.07 $ &  \\
         &  1183.032  &  -0.61  &  145875.697  & \nodata & \\
         &  1184.514  &  -0.21  &  145985.804  & \nodata &  \\
         &  1184.574  &  -0.92  &  145985.804  & \nodata &  \\
         &  1182.971  &  -0.92  &  145875.697  & $ -4.71 \pm 0.07 $ & STIS \\
         &  1183.032  &  -0.61  &  145875.697  & \nodata &  \\
         &  1184.514  &  -0.21  &  145985.804  & \nodata &  \\
         &  1184.574  &  -0.92  &  145985.804  & \nodata &  \\
         &  1324.314  &  -0.25  &  267238.390  & $ -4.84 \pm 0.14 $ &  \\
         &  1324.396  &  -0.10  &  267243.993  & \nodata &  \\
         &  1387.303  &  -0.32  &  245665.403  & $ -4.82 \pm 0.09 $ &  \\
         &  1387.382  &  -0.07  &  245701.305  & \nodata &  \\
         &  1387.994  &  -1.02  &  245701.305  & \nodata &  \\
         &  4103.390  &  -0.36  &  221302.200  & $ -4.97 \pm 0.03 $ & UVES \\
         &  4510.880  &  -0.46  &  287529.395  & $ -5.00 \pm 0.02 $ &  \\
         &  4510.960  &  -0.06  &  287591.492  & \nodata &  \\
         &  4514.850  &   0.22  &  287706.910  & $ -4.97 \pm 0.02 $ &  \\
\nfour   &  1036.119  &   0.78  &  420058.003  & $ -4.29 \pm 0.32 $ & FUSE \\
         &  1036.149  &   0.62  &  420049.575  & \nodata &  \\
         &  1036.196  &   0.45  &  420045.796  & \nodata &  \\
         &  1036.237  &  -0.28  &  420049.575  & \nodata &  \\
         &  1036.239  &  -0.28  &  420058.003  & \nodata &  \\
         &  1078.711  &   0.56  &  429159.583  & $ -4.25 \pm 0.61 $ &  \\
         &  1718.550  &  -0.29  &  130693.904  & $ -5.23 \pm 0.13$ & STIS \\
\nfive   &  1238.821  &  -0.51  &       0.000  & $ -4.08 \pm 0.26 $ & STIS \\
         &  1242.804  &  -0.81  &       0.000  & \nodata &  \\
\otwo    &  4069.623  &  0.15   &  206730.762  & $ -4.63 \pm 0.01 $ & UVES \\
         &  4069.882  &  0.34   &  206786.278  & \nodata &  \\
         &  4072.153  &  0.55   &  206877.870  & \nodata &  \\
         &  4075.862  &  0.69   &  207002.475  & \nodata &  \\
         &  4078.842  &  -0.28  &  206786.278  & \nodata &  \\
         &  4349.426  &  0.06   &  185499.122  & $ -4.41 \pm 0.01 $ &  \\
\othree  &  1138.535  &  -0.76  &  210461.787  & $ -4.11 \pm 0.17 $ &  FUSE \\
         &  1149.634  &  -1.08  &  197087.711  & $ -3.83 \pm 0.12 $ &  \\
         &  1150.884  &  -0.60  &  197087.711  & \nodata &  \\
         &  1153.775  &  -0.38  &  197087.711  & \nodata &  \\
         &  1660.809  &  -6.34  &  113.178     & $ -4.21 \pm 0.12 $ & STIS \\
         &  1666.150  &  -5.94  &  306.174     & \nodata &  \\
         &  3312.329  &  -0.64  &  267377.113  & $ -4.39 \pm 0.03 $ & UVES \\
         &  3340.765  &  -0.48  &  267633.992  & $ -4.33 \pm 0.02 $ &  \\
         &  3754.696  &  -0.10  &  267377.113  & $ -4.29 \pm 0.01 $ &  \\
         &  3757.232  &  -0.45  &  267258.719  & \nodata &  \\
         &  3759.875  &  0.16   &  267633.992  & \nodata &  \\
         &  3774.026  &  -0.60  &  267377.113  & $ -4.30 \pm 0.03 $ &  \\
         &  3791.275  &  -0.62  &  267633.992  & $ -4.37 \pm 0.02 $ &  \\
         &  3961.573  &  0.31   &  306586.073  & $ -4.77 \pm 0.03 $ &  \\
\ofour   &  1338.615  &  -0.63  &  180480.794  & $ -4.01 \pm 0.21 $ & STIS \\
         &  1342.990  &  -1.33  &  180724.206  & \nodata &  \\
         &  1343.514  &  -0.38  &  180724.206  & \nodata &  \\
         &  1399.780  &  -6.04  &  0.000       & $ -4.50 \pm 0.65 $ &  \\
         &  1401.157  &  -5.66  &  385.900     & $ -4.08 \pm 0.25 $ &  \\
         &  1407.382  &  -6.03  &  385.900     & \nodata &  \\
\ofive   &  1371.296  &  -0.33  &  158797.700  & $ -3.47 \pm 0.46 $ & STIS \\
\enddata
\tablecomments{Abundance relative to hydrogen: \abund{X}.}
\tablenotetext{a}{Multiplet.}
\end{deluxetable*}

\startlongtable
\begin{deluxetable*}{lcDrcc}
\tablecaption{Additional Absorption Features \label{tab:lines_other}}
\tablehead{
\colhead{Ion} & \colhead{$\lambda_{\rm lab}$} & \multicolumn2c{$\log gf$} & \colhead{$E_l$} & \colhead{Abundance} & \colhead{Instrument} \\
\colhead{} & \colhead{(\AA)} & \multicolumn2c{} & \colhead{(cm$^{-1}$)}
}
\decimals
\startdata
\netwo   &  3334.836  &  0.39  &  219130.777  & $ -5.39 \pm 0.02 $ & UVES \\
         &  3344.396  &  -0.27  &  219947.458  & $ -5.39 \pm 0.06 $ &  \\
         &  3355.017  &  0.14  &  219648.443  & $ -5.52 \pm 0.06 $ &  \\
         &  3664.073  &  -0.26  &  219130.777  & $ -5.72 \pm 0.04 $ &  \\
         &  3694.212  &  0.09  &  219130.777  & $ -5.86 \pm 0.02 $ &  \\
\mgtwo   &  4481.126  &  0.74  &  71490.190  & $ -6.05 \pm 0.06 $ & UVES \\
         &  4481.325  &  0.59  &  71491.064  & \nodata &  \\
\althree &  1605.766  &  0.26  &  53682.930  & $ -7.53 \pm 0.25 $ & STIS \\
         &  1854.716  &  0.06  &  0.000  & $ -7.40 \pm 0.28 $ & IUE \\
         &  1862.790  &  -0.24  &  0.000  & \nodata &  \\
\sithree &  994.790  &  -0.85  &  52853.283  & $ -5.65 \pm 0.14 $ & FUSE \\
         &  997.386  &  -0.62  &  53115.010  & \nodata &  \\
         &  1108.358  &  -0.06  &  52724.691  & $ -6.19 \pm 0.07 $ &  \\
         &  1109.940  &  -0.19  &  52853.283  & \nodata &  \\
         &  1109.970  &  0.29  &  52853.283  & \nodata &  \\
         &  1113.174  &  -1.36  &  53115.010  & $ -5.96 \pm 0.06 $ &  \\
         &  1113.204  &  -0.19  &  53115.010  & \nodata &  \\
         &  1113.230  &  0.56  &  53115.010  & \nodata &  \\
         &  1144.309  &  0.74  &  130100.523  & $ -5.65 \pm 0.22 $ &  \\
         &  1206.500  &  0.23  &  0.000  & $ -5.81 \pm 0.06 $ & STIS \\
         &  1206.555  &  0.73  &  82884.404  & \nodata &  \\
         &  1207.517  &  0.31  &  122214.527  & \nodata &  \\
         &  1294.545  &  -0.04  &  52853.283  & $ -6.09 \pm 0.03 $ &  \\
         &  1296.726  &  -0.13  &  52724.691  & \nodata &  \\
         &  1298.892  &  -0.26  &  52853.283  & \nodata &  \\
         &  1298.946  &  0.44  &  53115.010  & \nodata &  \\
         &  1301.149  &  -0.13  &  52853.283  & \nodata &  \\
         &  1303.323  &  -0.04  &  53115.010  & \nodata &  \\
\sifour  &  1066.614  &  0.72  &  160374.406  & $ -5.92 \pm 0.11 $ & FUSE \\
         &  1066.636  &  -0.59  &  160374.406  & \nodata &  \\
         &  1066.650  &  0.56  &  160375.589  & \nodata &  \\
         &  1128.325  &  -0.48  &  71748.643  & $ -5.98 \pm 0.12 $ &  \\
         &  1128.340  &  0.47  &  71748.643  & \nodata &  \\
         &  1393.755  &  0.03  &  0.000  & $ -5.92 \pm 0.08 $ & STIS \\
         &  1402.770  &  -0.28  &  0.000  & \nodata &  \\
         &  4088.862  &  0.20  &  193978.889  & $ -5.73 \pm 0.01$ & UVES \\
         &  4116.104  &  -0.11  &  193978.889  & \nodata &  \\
\pfour   &  950.657  &  0.27  &  0.000  & $ -8.24 \pm 0.04 $ & FUSE \\
         &  1028.094  &  -0.32  &  67918.033  & $ -8.55 \pm 0.11 $ &  \\
         &  1030.514  &  -0.44  &  68146.475  & \nodata &  \\
         &  1030.515  &  0.25  &  68615.174  & \nodata &  \\
         &  1033.112  &  -0.32  &  68146.475  & \nodata &  \\
         &  1035.516  &  -0.22  &  68615.174  & \nodata &  \\
\pfive   &  1117.977  &  -0.01  &  0.000  & $ -8.35 \pm 0.10 $ & FUSE \\
         &  1128.008  &  -0.32  &  0.000  & \nodata &  \\
\sthree  &  1015.554  &  -1.83  &  297.200  & $ -5.88 \pm 0.10 $ & FUSE \\
         &  1015.554  &  -1.86  &  297.200  & \nodata &  \\
         &  1015.787  &  -1.78  &  297.200  & \nodata &  \\
         &  1077.134  &  -1.08  &  11320.000  & $ -5.92 \pm 0.11 $ &  \\
         &  1190.208  &  -1.66  &  0.000  & $ -6.26 \pm 0.18 $ &  STIS \\
         &  1194.041  &  -1.31  &  297.200  & $ -6.30 \pm 0.08 $ &  \\
         &  1194.433  &  -1.79  &  297.200  & \nodata &  \\
         &  1200.956  &  -1.03  &  832.500  & $ -6.23 \pm 0.08 $ &  \\
\sfour   &  999.787  &  -1.06  &  133619.597  & $ -6.16 \pm 0.18 $ & FUSE \\
         &  1006.097  &  -1.76  &  134245.405  & $ -6.41 \pm 0.21 $ &  \\
         &  1006.403  &  -0.81  &  134245.405  & \nodata &  \\
         &  1062.678  &  -1.09  &  0.000  & $ -6.33 \pm 0.16 $ &  \\
         &  1072.996  &  -0.83  &  951.100  & $ -6.49 \pm 0.17 $ &  \\
         &  1073.528  &  -1.79  &  951.100  & \nodata &  \\
         &  1098.357  &  -1.75  &  94103.097  & $ -6.43 \pm 0.13 $ &  \\
         &  1098.917  &  -0.61  &  94150.403  & \nodata &  \\
         &  1099.472  &  -0.80  &  94103.097  & \nodata &  \\
         &  1100.040  &  -1.75  &  94150.403  & \nodata &  \\
         &  1138.076  &  -1.79  &  123509.301  & $ -6.08 \pm 0.29 $ &  \\
         &  1138.210  &  -1.49  &  123509.301  & \nodata &  \\
\sfive   &  1039.917  &  0.24  &  270700.417  & $ -5.79 \pm 0.34 $ & FUSE \\
         &  1122.042  &  0.09  &  234956.003  & $ -6.01 \pm 0.27 $ &  \\
         &  1128.667  &  -0.07  &  234947.098  & \nodata &  \\
         &  1128.776  &  -0.97  &  234956.003  & \nodata &  \\
         &  1133.902  &  -0.24  &  234941.494  & \nodata &  \\
         &  1133.973  &  -0.97  &  234947.098  & \nodata &  \\
         &  1501.760  &  -0.50  &  127150.700  & $ -6.71 \pm 0.17 $ & STIS \\
\ssix    &  944.523  &  -0.35  &  0.000  & $ -5.51 \pm 0.26 $ & FUSE \\
\clfour  &  977.890  &  -0.39  &  491.000  & $ -9.30 \pm 0.15 $ & FUSE \\
         &  984.950  &  0.36  &  1341.000  & $ -9.08 \pm 0.14 $ &  \\
\fethree &  981.369  &  0.06  &  20051.100  & $ -5.92 \pm 0.31 $ & FUSE \\
         &  983.860  &  -0.03  &  20300.801  & $ -5.89 \pm 0.29 $ &  \\
         &  1124.881  &  -0.44  &  436.200  & $ -6.40 \pm 0.65 $ &  \\
\fefour  &  1661.573  &  0.47  &  190318.345  & $ -6.11 \pm 0.10 $ & STIS \\
         &  1662.319  &  -0.15  &  127929.122  & \nodata &  \\
         &  1662.519  &  0.29  &  160311.636  & \nodata &  \\
         &  1663.543  &  -0.33  &  127766.148  & \nodata &  \\
         &  1668.065  &  -0.85  &  127929.122  & $ -6.13 \pm 0.10 $ &  \\
         &  1668.119  &  -0.57  &  167712.501  & \nodata &  \\
         &  1668.177  &  -0.98  &  162074.425  & \nodata &  \\
         &  1668.422  &  -1.21  &  128967.669  & \nodata &  \\
         &  1669.602  &  -0.79  &  128191.539  & \nodata &  \\
         &  1669.666  &  -2.71  &  167712.501  & \nodata &  \\
         &  1669.813  &  -0.90  &  128541.855  & \nodata &  \\
         &  1690.007  &  -0.04  &  156049.316  & $ -6.48 \pm 0.13 $ &  \\
         &  1690.305  &  -0.36  &  153651.737  & \nodata &  \\
         &  1690.321  &  -0.30  &  156224.876  & \nodata &  \\
         &  1690.577  &  -1.22  &  162087.815  & \nodata &  \\
         &  1690.635  &  -0.48  &  159010.389  & \nodata &  \\
         &  1690.697  &  -2.21  &  159227.901  & \nodata &  \\
         &  1690.764  &  -1.29  &  162074.425  & \nodata &  \\
\nifour  &  1432.449  &  -0.12  &  121807.707  & $ -7.36 \pm 0.18 $ & STIS \\
         &  1433.108  &  0.17  &  147635.901  & \nodata &  \\
         &  1433.207  &  -1.61  &  145962.499  & \nodata &  \\
         &  1433.234  &  -0.89  &  144815.093  & \nodata &  \\
         &  1433.260  &  -0.03  &  159498.495  & \nodata &  \\
         &  1444.143  &  -0.62  &  139886.707  & $ -7.19 \pm 0.20 $ &  \\
         &  1444.417  &  -0.39  &  122386.102  & \nodata &  \\
         &  1444.913  &  0.30  &  145702.200  & \nodata &  \\
         &  1445.078  &  0.56  &  190864.703  & \nodata &  \\
         &  1445.078  &  0.56  &  179655.002  & \nodata &  \\
         &  1445.179  &  -0.45  &  141831.995  & \nodata &  \\
         &  1481.411  &  -0.25  &  140343.004  & $ -6.94 \pm 0.15 $ &  \\
         &  1482.248  &  0.33  &  139289.405  & \nodata &  \\
         &  1482.551  &  -0.69  &  139289.405  & \nodata &  \\
         &  1482.665  &  -0.02  &  145962.499  & \nodata &  \\
         &  1534.710  &  0.44  &  110410.598  & $ -6.96 \pm 0.11 $ &  \\
         &  1534.931  &  -0.37  &  121807.707  & \nodata &  \\
         &  1536.842  &  -0.29  &  141577.202  & \nodata &  \\
         &  1537.187  &  -0.68  &  148358.199  & \nodata &  \\
         &  1537.248  &  0.01  &  111195.796  & \nodata &  \\
\gefour  &  1189.028  &   0.01  &       0.000  & $ -10.75 \pm 0.11 $ & STIS \\
         &  1229.840  &  -0.30  &       0.000  & $ -10.92 \pm 0.14 $ &  \\
\enddata
\tablecomments{Abundance relative to hydrogen: \abund{X}.}
\end{deluxetable*}

\clearpage





\end{document}